\documentclass[aps, twocolumn, superscriptaddress, showpacs, longbibliography]{revtex4-1}

\usepackage{amsmath, amsfonts, amssymb, amsthm}
\usepackage{hyperref}
\usepackage{dsfont}
\usepackage{physics}
\usepackage{tikz}
\usepackage{usefulnotations}

\begin{document}
	\title{Quantum State Evolution and Berry Potentials at Exceptional Points and Quantum Phase Transitions}

	\author{Chia-Yi Ju}
	\email{chiayiju@mail.nsysu.edu.tw}
	\affiliation{Department of Physics, National Sun Yat-sen University, Kaohsiung 80424, Taiwan}
	\affiliation{Center for Theoretical and Computational Physics, National Sun Yat-sen University, Kaohsiung 80424, Taiwan}
	\affiliation{Physics Division, National Center for Theoretical Sciences, Taipei 106319, Taiwan}
	\author{Fu-Hsiang Huang}
	\affiliation{Department of Physics, National Taiwan University, Taipei 10617, Taiwan}

	\begin{abstract}
		The behavior of quantum states at exceptional points and at critical points associated with quantum phase transitions is intriguing yet puzzling. In this study, we present an alternative method for obtaining the Berry potentials using the evolution generator along the parameter induced dimension and demonstrate that they are singular at these critical points. Although these singularities may appear to indicate a breakdown in quantum state evolution, we show that the information carried by quantum states evolving across these critical points is not destroyed. Specifically, when the evolution generator of the full Hilbert space bundle is taken into account, the quantum states remain insensitive to the critical points. In physical terms, it's similar to the classical image of an object smoothly passing through a black hole's event horizon. Further similarities between exceptional points and quantum phase transitions are explored in this work.
	\end{abstract}

	\maketitle

	\section{Introduction}

		It is well known that the ability of quantum states to carry information is crucial for quantum computing.

		In principle, the evolution of the quantum states is reversible; therefore, information loss in a closed system seems impossible. Nevertheless, the information carried by a quantum state can be reduced at the point where two or more Hamiltonian eigenstates merge into one, leading to a loss of degrees of freedom in the eigenbasis. Since these points, known as exceptional points (EPs)~\cite{Kato1976, Heiss2004, Ozdemir2019}, are often associated with quantum phase transitions (QPTs)~\cite{Heiss1998}, they share some properties.

		Therefore, we study the aforementioned information paradox~\cite{Znojil2020, Znojil2021} by investigating how quantum states evolve across an EP and the critical point of a QPT.

		Typically, Berry potentials (often called Berry connections)\cite{Berry1984, Xiao2010, Sakurai2017} are used to study the parameter dependence of quantum states under the adiabatic approximation, where an eigenstate evolves into the eigenstate with the closest eigenvalue. Specifically, the Berry potentials describe how eigenstates evolve with respect to the parameter (see Appendix~\ref{App:BerryPotential} for a brief review).

		However, at an EP, the Hamiltonian's eigenstates do not form a complete basis, causing the state evolution across the EP to undergo rapid changes. Similarly, during a QPT, the ground states are degenerate, leading to ambiguity about which state the ``ground state'' evolves into.

		Here, we present a geometric approach to study the parameter dependence of quantum states, applicable to both the Hermitian and non-Hermitian regimes. Additionally, since quantum states can change rapidly at certain critical points, the assumption of the adiabatic approximation may not be appropriate near the critical point of a QPT, let alone at an EP where some eigenstates coalesce.

		Inspired by $\cal{PT}$-symmetric quantum mechanics~\cite{Bender1998, Bender2004, Bender2007} and pseudo-Hermitian quantum mechanics~\cite{Mostafazadeh2003, Mostafazadeh2010}, studies have shown that the Hilbert space can be dynamic~\cite{Mostafazadeh2004, Mostafazadeh2018, Ju2019, Mostafazadeh2020, Ju2022, Mostafazadeh2024, Ju2025}.

		Moreover, a recent study~\cite{Ju2024} shows that time is not the only direction in which a state can evolve; when the Hamiltonian of the quantum system, whether time-dependent~\cite{Znojil2024} or independent, depends on a continuous parameter, the parameter induces an additional evolution dimension.

		As will be shown shortly, the governing equation for state evolution in the parameter-induced evolution dimension (or emergent dimension, for short) is a Schr\"{o}dinger-like equation where the Hamiltonian (the generator in the time dimension) is replaced by an evolution generator in the parameter space. Roughly speaking, the generators in time (i.e. the Hamiltonian) and the emergent dimension are fiber bundle version of the Christoffel symbols in (pseudo-)Riemannian geometry~\cite{Nakahara2003}. 

		Nevertheless, like the Christoffel symbols, the generator in the emergent dimension is gauge-dependent. We therefore introduce a gauge that is compatible with the adiabatic theorem~\cite{Born1928, Kato1950}, so that the Berry potential~ can be readily obtained by projecting the evolution generator onto the eigenstates of the Hamiltonian. The adiabatic-theorem-compatible gauge is also introduced in the following section.

		Interestingly, based on experience, the Berry potentials are typically singular when the Hamiltonian is at an EPs{and when the eigenvalues of the Hamiltonian degenerate (diabolic points, or DPs). In other words, the evolution in the emergent dimension can sometimes encounter singularities.

		Therefore, this study investigates the singularities at the EPs and DPs, motivated by their physical significants. For example, they are often indicators of the QPTs~\cite{Kato1976, Zanardi2006, Cejnar2007, Sachdev2009, SHIJIAN2010, Cejnar2006, PerezFernandez2009, Tzeng2017, Znojil2020, Znojil2021, Tzeng2021, Wu2023}.

		In the following, we review the emergence of the parameter-induced dimension, the generator, and the adiabatic gauge (namely, the gauge compatible with the adiabatic theorem), as well as its relation to the Berry potential. We then analyze the mathematical structures of the singularities and conlude by discussing the physical implications of the results.

	\section{Background Review}

		In this section, we review some ideas that will be beneficial for the discussions that follow.

		\subsection{Emergent Dimension}

			The evolution of a quantum state in time $t$ is governed by the Schr\"{o}dinger equation, which can be viewed as a parallel transport in a Hilbert space bundle~\cite{Ju2019}, i.e.,
			\begin{align}
				0 = \nabla_t \ket{\psi} \equiv \left( \partial_t + i H \right) \ket{\psi}, \label{SchrodingerEq}
			\end{align}
			where $\nabla_t$ represents the covariant derivative or the connection in the $t$-direction and $H$ is the Hamiltonian. By attributing such a geometrical interpretation to the Schr\"{o}dinger equation, we can naturally generalize the conventional quantum mechanical inner product, i.e., $\braket{\phi}{\psi}$, to a connection-compatible inner product
			\begin{align}
				\Braket{\phi}{\psi} = \bra{\phi} G \ket{\psi},
			\end{align}
			where $\Ket{\psi} = \ket{\psi}$, $\Bra{\phi} = \bra{\phi} G$, and $G$ is the Hilbert space metric governed by the equation
			\begin{align}
				0 = \nabla_t G = \partial_t G - i G H + i H^\dagger G.
			\end{align}

			Interestingly, it was found that if the Hamiltonian is a function of a continuous parameter $q$, the parameter $q$ induces an emergent dimension~\cite{Ju2024} of the Hilbert space in addition to $t$. Specifically, the state evolving in the $q$-direction is described by a Schr\"{o}dinger-like equation,
			\begin{align}
				0 = \nabla_q \ket{\psi} = \left(\partial_q + i K\right) \ket{\psi}, \label{qEvolution}
			\end{align}
			where $K = K(t, q)$ is the evolution generator in the $q$-direction.

			To determine the $q$-evolution generator $K$, we note that the commutator between $\nabla_t$ and $\nabla_q$ acting on any quantum state is zero. To be more specific,
			\begin{align}
				\left[\nabla_t, \nabla_q\right] \ket{\psi} & = \left(\nabla_t \nabla_q - \nabla_q \nabla_t\right) \ket{\psi}\\
				& = \nabla_t \left(\nabla_q \ket{\psi}\right) - \nabla_q \left(\nabla_t \ket{\psi}\right) = 0,
			\end{align}
			where the last equality follows from Eqs.~\eqref{SchrodingerEq} and \eqref{qEvolution}.

			On the other hand, if we work out the commutator between $\nabla_t$ and $\nabla_q$, we find
			\begin{align}
				\left[\nabla_t, \nabla_q\right] \ket{\psi} = \left(\partial_t K - i \left[ K, H \right] - \partial_q H\right) \ket{\psi}.
			\end{align}

			Thus, the $q$-evolution generator $K$ obeys
			\begin{align}
				\partial_t K - i \left[ K, H \right] - \partial_q H = 0. \label{MainEquation}
			\end{align}

			Interestingly, the fact that the vanishing commutator $\left[\nabla_t, \nabla_q\right]$ not only allows us to determine $K$, but also implies that the local curvature $\mathcal{F}$ two-form vanishes everywhere. In particular, from the definition of the local curvature two-form $\mathcal{F}$,
			\begin{align}
				\mathcal{F} = \dfrac{1}{2}\left(F_{tq} dt \wedge dq + F_{qt} dq \wedge dt\right),
			\end{align}
			where $F_{tq} \ket{\psi} = - F_{qt} \ket{\psi} = - i \left[\nabla_t, \nabla_q\right] \ket{\psi} = 0$, it follows that
			\begin{align}
				\mathcal{F} = 0. \label{LocalCurvatureTwoForm}
			\end{align}

			In other words, it is locally flat eveywhere.

			Besides the states, the metric can also evolve in the $q$-direction, governed by
			\begin{align}
				0 = \nabla_q G = \partial_q G - i G K + i K^\dagger G. \label{Metricinq}
			\end{align}

			For the rest of this paper, we focus on the case where $H$ is time-independent but depends on a continuous parameter $q$, i.e., $H = H (q)$.

		\subsection{Adiabatic Gauge}

			Since the governing equation for $K$ [i.e., Eq.~\eqref{MainEquation}] is a differential equation, $K$ is obviously not unique because we can arbitrarily add to $K$ a $\Delta \! K$ term that satisfies
			\begin{align}
				\partial_t \Delta \! K - i \left[ \Delta \! K, H \right] = 0, \label{GaugeTransform}
			\end{align}
			so that $K' = K + \Delta \! K$ is also a solution to Eq.~\eqref{MainEquation}. In fact, the freedom of adding $\Delta \! K$ to $K$ comes from the gauge freedom associated with the Hilbert space metric $G$~\cite{Ju2019, Ju2024}.

			Analogous to the gauge redundancy in electrodynamics, we can take advantage of the gauge degree of freedom to choose a gauge (e.g., the Coulumb gauge and Lorenz gauge in electrodynamics) that is convenient to work with. For many purposes, such as working in biorthogonal quantum mechanics~\cite{Brody2013}, studying QPTs (fidelity susceptibility), and calculating Berry phases, it is often more convenient to choose a gauge in which the Hamiltonian eigenstates, evolving in the $q$-direction, remain eigenstates of the Hamiltonian, i.e.,
			\begin{align}
				H(q) \ket{\psi_{n} (t, q)} = h_{n} (q) \ket{\psi_{n} (t, q)}, \label{EigenstateChoice}
			\end{align}
			for all Hamiltonian eigenstates $\ket{\psi_{n}}$ within some range of $q$. Note that the condition shown in Eq. (3) does not always hold, as it assumes that each $\ket{\psi_{n} (t, q)}$ is a well-defined function of $q$, which is not always true (e.g., when an EP exists). This condition is motivated by the adiabatic theorem discussed in Appendix~\ref{App:Adiabatic}.

			To find a gauge condition compatible with Eqs.~\eqref{qEvolution} and \eqref{EigenstateChoice}, we take a $q$-derivative on both sides of the equation, along with Eqs.~\eqref{SchrodingerEq} and \eqref{qEvolution} and find that
			\begin{align}
				& \partial_q \left( H \ket{\psi_n} \right) = \partial_q \left( h_n \ket{\psi_n} \right)\\
				\Rightarrow & \left( \partial_t K \right) \ket{\psi_n} = \left( \partial_q h_n \right) \ket{\psi_n}. \label{AdiabaticConsequence}
			\end{align}
			The complete derivations are given in the Appendix of Ref.~\cite{Ju2024}.

			In other words, Eq.~\eqref{EigenstateChoice} (the main result in the adiabatic theorem) implies that $\partial_t K$ shares the same eigenstates with $H$, which is generally equivalent to
			\begin{align}
				\left[ \partial_t K , H \right] = 0; \label{Adiabatic}
			\end{align}
			this condition is therefore called the adiabatic gauge. In fact, the adiabatic gauge shown in Eq.~\eqref{Adiabatic} leads to Eq.~\eqref{AdiabaticConsequence} (see Appendix~\ref{App:Adiabatic} for a detailed derivation).

			Besides ensuring that the state remains an eigenstate as the parameter $q$ varies, the gauge choice in Eq.~\eqref{Adiabatic} restricts the evolution generator $K$ to be at most linear in $t$ for time-independent Hamiltonians. It can be easily demonstrated by taking a $t$-derivative on both sides of Eq.~\eqref{MainEquation}, i.e.,
			\begin{align}
				0 & = \partial_t^2 K - i \partial_t \left[ K , H \right] - \partial_t \partial_q H = \partial_t^2 K,
			\end{align}

			where we have used the fact that $H$ is time-independent (i.e., $\partial_t H = 0$) and Eq.~\eqref{Adiabatic}. Therefore, the evolution generator $K$ can be decomposed as
			\begin{align}
				K = K^{(1)} t + K^{(0)},
			\end{align}
			where $K^{(1)}$ and $K^{(0)}$ are time-independent and satisfy
			\begin{align}
				& K^{(1)} = i \left[ K^{(0)} , H\right] + \partial_q H,\\
				& \left[ K^{(1)} , H\right] = 0. \label{K1HEquals0}
			\end{align}

			Note that this gauge fixing condition does not fix all the gauge freedom. The residual gauge transformation, $K' = K + \Delta \! K$, under the adiabatic gauge is
			\begin{align}
				& K^{\prime(1)} = K^{(1)} + \Delta \! K^{(1)}\\
				& K^{\prime(0)} = K^{(0)} + \Delta \! K^{(0)}.
			\end{align}
			Then the gauge transformation in Eq.~\eqref{GaugeTransform} is reduced to
			\begin{align}
				\Delta \! K^{(1)} = i \left[ \Delta \! K^{(0)}, H \right].
			\end{align}
			Due to Eq.~\eqref{K1HEquals0}, $\Delta \! K^{(1)}$ and $\left[ \Delta \! K^{(0)}, H \right]$ are linear independent, since $\Delta \! K^{(1)}$ consists of the zero modes of $[\cdot, H]$ while $\left[ \Delta \! K^{(0)}, H \right]$ consists of the non-zero modes. Therefore, the residual gauge transformation gauge becomes
			\begin{align}
				\Delta \! K^{(1)} = 0 \quad \text{and} \quad \left[ \Delta \! K^{(0)}, H \right] = 0.
			\end{align} Consequently, the residual gauge freedom left is
			\begin{align}
				K' = K + \Delta \! K, \label{ResidualGaugeTransformation}
			\end{align}
			where $[\Delta \! K, H] = 0$. That is, the time dependent part in $K$ is uniquely determined in the adiabatic gauge.

			In fact, Eq.~\eqref{ResidualGaugeTransformation} corresponds to a difference choice of eigenstates in Eq.~\eqref{EigenstateChoice}, i.e.,
			\begin{align}
				H(q) \ket{\widetilde{\psi}_{n} (t, q)} = h_{n} (q) \ket{\widetilde{\psi}_{n} (t, q)},
			\end{align}
			where
			\begin{align}
				\ket{\widetilde{\psi}_{n}(t, q)} = e^{-i \chi_{n} (q)}\ket{\psi_{n}(t, q)},
			\end{align}
			such that
			\begin{align}
				& \partial_q \ket{\widetilde{\psi}_i(t, q)} = - i K' \ket{\widetilde{\psi}_n(t, q)}\\
				& = - i (K + \Delta \! K) \ket{\widetilde{\psi}_n(t, q)}\\
				& = e^{-i \chi_{n} (q)}\left[ \partial_q \chi_{n} (q)\right] \ket{\psi_{n} (t, q)} + e^{-i \chi_{n} (q)} \partial_q \ket{\psi_{n}(t, q)}\\
				& = e^{-i \chi_{n} (q)}\left[ \partial_q \chi_{n}(q)\right] \ket{\psi_{n}(t, q)} - i e^{-i \chi_{n} (q)} K \ket{\psi_{n}(t, q)}\\
				& = \left[- i K + \partial_q \chi_{n}(q)\right] \ket{\widetilde{\psi}_{n}(t, q)}.
			\end{align}

			In other words,
			\begin{align}
				& \Delta \! K \ket{\widetilde{\psi}_n(t, q)} = \partial_q \chi_{n}(q) \ket{\widetilde{\psi}_{n}(t, q)} \label{DeltaKEigenvalue}\\
				\Rightarrow & [\Delta \! K, H] = 0.
			\end{align}

			Note that the reason why the gauge degrees of freedom do not affect the time-dependent part can also be explained in the same manner~\cite{Chen2025}.

		\subsection{Berry Potentials from Evolution Generator \label{Sec:Berry}}

			Besides making $K$ neater and simpler for calculation, the adiabatic gauge is also essential for relating $K$ to the Berry potential, a well-developed tool for investigating and classifying many interesting physical phenomena.

			Therefore, in this subsection, we show that the Berry potential---i.e., the function $A_i(q)$ defined as
			\begin{align}
				A_{n}(q) = i \Bra{\phi_{n}(q)} \partial_q \Ket{\phi_{n}(q)},
			\end{align}
			where $\ket{\phi_{n}(q)}$ is an eigenstate of the Hamiltonian $H(q)$---is, in fact, a projection of the $q$-evolution generator $K$, in the adiabatic gauge, onto a specific eigenstate. Additionally, we relate the residual gauge freedom to the gauge freedom in the Berry potential.

			Specifically, given $\Braket{\phi_{m} (q)}{\phi_{n} (q)} = \delta_{{mn}}$ (i.e., choosing the metric $G$ such that $\Bra{\phi_{n}} = \bra{\phi_{n}} G$ is a left eigenstate of the Hamiltonian~\cite{Tzeng2021}), the Berry potential can be obtained as
			\begin{align}
				A_{n}(q) = \Bra{\phi_{n}(q)} K(t_0, q) \Ket{\phi_{n}(q)},
			\end{align}
			for any $t = t_0$. Appendix ~\ref{App:ExampleKToA} presents a non-Hermitian example, taken from \cite{MehriDehnavi2008}, for which we reproduce the results using our new method. (A Hermitian example can be found in \cite{Ju2024}, where the Berry curvature is also calculated.) As we will see shortly, the choice of $t_0$ is equivalent to a gauge choice.

			Since the Berry potential is derived under the adiabatic approximation, i.e.,
			\begin{align}
				H(q) \ket{\phi_{n}(q)} = h_{n}(q) \ket{\phi_{n}(q)},
			\end{align}
			which is a special case of Eq.~\eqref{EigenstateChoice} by choosing $\ket{\psi_{n}(t = 0, q)} = \ket{\phi_{n}(q)}$. With $\Braket{\phi_{m} (q)}{ \phi_{n} (q)} = \delta_{{mn}}$, we find
			\begin{align}
				A_{n}(q) & = i \Bra{\phi_{n} (q)} \left(\partial_q \Ket{\phi_{n} (q)}\right)\\
				& = i \Bra{\psi_{n} (0, q)} \left(\partial_q \Ket{\psi_{n} (0, q)}\right)\\
				& = \Bra{\psi_{n} (0, q)} K(t = 0, q) \Ket{\psi_{n} (0, q)}\\
				& = \Bra{\psi_{n} (0, q)} K^{(0)} \Ket{\psi_{n} (0, q)}.
			\end{align}

			In fact, adding the residual gauge freedom to $K$, i.e., $K' = K + \Delta \! K$, yields the standard Berry potential gauge transformation
			\begin{align}
				A'_{n}(q) = A_{n}(q) + \partial_q \chi_n(q),
			\end{align}
			where $\partial_q \chi_n(q) = \Bra{\psi_{n} (0, q)} \Delta \! K \Ket{\psi_{n} (0, q)}$. A detailed derivation can be found in Appendix~\ref{App:GaugeFreedoms}.

			Therefore, the Berry potential of an eigenstate can be determined by the $q$-evolution generator $K$ in the adiabatic gauge.

	\section{The Singularities}

		We next present two analytically solvable examples: one at an EP and the other at a DP, both exhibiting singular behavior under the adiabatic gauge, which does not appear in other gauge choices. Detailed mathematical analyses of the occurrence and its avoidance are provided in Appendices~\ref{App:EP} and \ref{App:DP}.

		\subsection{EP Sigularities}

			Although the adiabatic gauge is physically more compelling because it is related to the Berry potentials and reduces the differential equations to algebraic equations, the evolution generator $K$ can sometimes be singular when the adiabatic gauge fixing is applied, even if the $H(q)$ is well-behaved for every $q$. This is reflected in the singular behavior of the Berry potential (see Appendix~\ref{App:ExampleKToA} for an example).

			As a toy example, we consider a dimensionless Hamiltonian
			\begin{align}
				H_{\text{\tiny EP}}( q ) = \begin{pmatrix}
					i q & 1\\
					1 & -i q
				\end{pmatrix}, \label{EPHamiltonain}
			\end{align}
			with $q$ being the parameter which induces the emergent evolution dimension. Applying Eqs.~\eqref{MainEquation} and \eqref{Adiabatic}, we find that the $q$-direction generator $K_{\text{\tiny EP}}$ is given by
			\begin{align}
				K_{\text{\tiny EP}} = \begin{pmatrix}
					\dfrac{i q^2 t}{q^2 - 1} + \alpha_1 + i q \alpha_2 & \dfrac{2 q t + 1}{2\left(q^2 - 1\right)} + \alpha_2\\
					\dfrac{2 q t - 1}{2\left(q^2 - 1\right)} + \alpha_2 & \dfrac{-i q^2 t}{q^2 - 1} + \alpha_1 - i q \alpha_2
				\end{pmatrix}, \label{EPSolution}
			\end{align}
			where $\alpha_1$ and $\alpha_2$ are two undetermined functions of $z$, corresponding to the residual gauge freedoms. It is obvious that there are no $\alpha$'s that remove the singularity at $q = \pm 1$, which are, in fact, the two EPs in the Hamiltonian.

			In other words, since the evolution of the state is governed by Eq.~\eqref{qEvolution}, the state changes too fast when the parameter goes across an EP. Specifically, the $K$ in
			\begin{align}
				\ket{\psi (t, q + \varepsilon)} \approx \ket{\psi (t, q)} - \varepsilon i K (t, q) \ket{\psi (t, q)}
			\end{align}
			diverges at $q \rightarrow \pm 1$. Hence, the quantum state seems unable to evolve across the EP, erasing the information it carries.

			These singularities at the EPs are, in fact, expected. Specifically, although the adiabatic gauge can be applied almost everywhere, even when it is arbitrarily close to an EP, the eigenstates of $H$ do not form a complete basis at the EPs; hence, $H$ and $\partial_t K$ do not share the same set of eigenstates in general.

			Nevertheless, since the local curvature two-form vanishes everywhere [Eq.~\eqref{LocalCurvatureTwoForm}], we should expect the states to vary smoothly with respect to $q$ in general, even at an EP. Therefore, the singularity is an ``artifact'' arising from the choice of gauge. In other words, the singularities are gauge singularities, analogous to the coordinate singularity at a classical black hole's event horizon.

			Here, we provide another evolution generator $K'_{\text{\tiny EP}}$ for the Hamiltonian in Eq.~\eqref{EPHamiltonain} without applying the adiabatic gauge:
			\begin{align}
				K'_{\text{\tiny EP}} = \left\lbrace\begin{array}{ll}
					\dfrac{1}{3}\begin{pmatrix}
						- 2 i t^3 + 3 i t & - 2 t^3 - 3 t^2\\
						- 2 t^3 + 3 t^2 & 2 i t^3 - 3 i t
					\end{pmatrix}, & \text{for $q = 1$}\\
					\dfrac{1}{3}\begin{pmatrix}
						- 2 i t^3 + 3 i t & 2 t^3 - 3 t^2\\
						2 t^3 + 3 t^2 & 2 i t^3 - 3 i t
					\end{pmatrix}, & \text{for $q = -1$}\\
					\dfrac{1}{2 \zeta^3}\begin{pmatrix}
						k_{11} & k_{12}\\
						k_{21} & k_{22}
					\end{pmatrix}, & \text{otherwise}
				\end{array}\right.,
			\end{align}
			where
			\begin{align}
				k_{11} & = i \sin(2 \zeta t) - 2 i \zeta q^2 t = k_{22}^*,\\
				k_{12} & = q \sin(2 \zeta t) + \zeta\left[ \cos (2 \zeta t) - 2 q t - 1 \right],\\
				k_{21} & = q \sin(2 \zeta t) - \zeta\left[ \cos (2 \zeta t) + 2 q t - 1 \right],
			\end{align}
			and $\zeta = \sqrt{1 - q^2}$, so that the evolution generator is continuous everywhere including at $q = \pm 1$. This means that the singularities can be removed by another choice of gauge.

			In other words, the singularities at the EPs only arise when we use the Hamiltonian's eigenstates as the basis. However, this singularity can be avoided by choosing a different gauge. Therefore, the information carried by the state remains intact.

			A general discussion on how to remove the singularities at an EP can be found in Appendix~\ref{App:EP}.

		\subsection{DP Singularities}

			Interestingly, the singularities not only occur at the EPs but sometimes appear when the Hamiltonian is at a DP, even though the eigenstates still form a complete set. We will show in the next section that this phenomenon is closely related to QPTs.

			As an example, we consider a periodic Su-Schrieffer-Heeger (SSH) model~\cite{Su1979}, namely, a one-dimensional fermionic chain with alternating coupling constants between sites (see Fig.~\ref{Fig:SSH}).

			The Hamiltonian of the system can be expressed as
			\begin{align}
				H_{\text{\tiny SSH}} = \sum_{n = 0}^{N - 1} \left(v a^\dagger_{n} b_{n} + w a^\dagger_{n + 1} b_{n}\right) + \text{h.c.},
			\end{align}
			where $a_n$ ($a_n^\dagger$) and $b_n$ ($b_n^\dagger$) are the annihilation (creation) operators for the A and B sites, respectively, in the $n$th cell; $v$ and $w$ are the coupling constants between sites; $N$ is the number of unit cells; and $a_N = a_0$, $b_N = b_0$.

			It is well known that the ground state and the first excited state undergo a level crossing at $v = w$ in the thermodynamic limit. In other words, the system reaches a DP at $v = w$ when $N \rightarrow \infty$.

			\begin{figure}[h!]
				\begin{tikzpicture}
					\foreach \i in {0, 2.5, 5}{
						\begin{scope}[shift = {({-0.5 + \i}, 0)}]
							\shade[ball color = red, shading = ball, opacity = 1] (0, 0) circle (0.3);
							\node[color = white] at (0, -0.1) {\small A};
						\end{scope}
						\begin{scope}[shift = {({0.5 + \i}, 0)}]
							\shade[ball color = blue, shading = ball, opacity = 1] (0, 0) circle (0.3);
							\node[color = white] at (0, -0.1) {\small B};
						\end{scope}

						\draw[<->, color = green!50!black] ({-0.3 + \i}, 0.4) arc (120: 60: 0.6) node [midway, above] {$v$};
					}
					\draw (0, -0.5) arc (190: 230: 0.5) node [below right] {$n - 1$};
					\draw (2.5, -0.5) arc (190: 230: 0.5) node [below right] {$n$};
					\draw (5, -0.5) arc (190: 230: 0.5) node [below right] {$n + 1$};

					\foreach \i in {-1, 1.5, 4}{
						\draw [densely dashed, color = gray] (\i, 0.5) -- (\i, -0.5);
					}

					\foreach \i in {-1, 1.5}{
						\draw[<->, color = orange!70!black, line width = 1] ({1.7 + \i}, 0.4) arc (120: 60: 1) node [midway, above] {$w$};
					}
					\node at (- 1.4, 0) {\Large $\cdots$};
					\node at (6.4, 0) {\Large $\cdots$};
				\end{tikzpicture}
				\caption{Illustration of cells numbered $n - 1$, $n$, and $n + 1$. The red balls represent the A sites, and the blue balls represent the B sites within each cell. The coupling between A and B sites within the same cell is denoted by $v$, while the coupling between neighboring cells is denoted by $w$.}
				\label{Fig:SSH}
			\end{figure}

			For simplicity, we rescale the Hamiltonian by $1/v$ so that the coupling constant within a unit cell is \emph{one} in the rescaled Hamiltonian, i.e.,
			\begin{align}
				H_{\text{\tiny rSSH}} = \sum_{n = 0}^{N - 1} \left(a^\dagger_{n} b_{n} + g a^\dagger_{n + 1} b_{n}\right) + \text{h.c.}, \label{HSSH}
			\end{align}
			where $g = w / v$ is the relative coupling constant. This means that the DP in the rescaled Hamiltonian corresponds to $g = 1$.

			We collect and redefine (Fourier transform) the lowering and raising operators as
			\begin{align}
				& \left\lbrace \begin{array}{l}
					\displaystyle C_{\text{A}k} = \frac{1}{\sqrt{N}} \sum_{n = 0}^{N - 1} e^{i n \theta_k} a_n\\
					\displaystyle C_{\text{B}k} = \frac{1}{\sqrt{N}} \sum_{n = 0}^{N - 1} e^{i n \theta_k} b_n
				\end{array} \right.\\
				& \Leftrightarrow \left\lbrace \begin{array}{l}
					\displaystyle C_{\text{A}k}^\dagger = \frac{1}{\sqrt{N}} \sum_{n = 0}^{N - 1} e^{- i n \theta_k} a_n^\dagger\\
					\displaystyle C_{\text{B}k}^\dagger = \frac{1}{\sqrt{N}} \sum_{n = 0}^{N - 1} e^{- i n \theta_k} b_n^\dagger
				\end{array} \right.,
			\end{align}
			where $\theta_k = 2 k \pi / N$ and $k$ is an integer, such that
			\begin{align}
				\left\{C_{\text{M}k}, C_{\text{N}k'}^\dagger\right\} = \delta_{\text{MN}} \delta_{k k'}.
			\end{align}
			Moreover, the Hamiltonian is simplified to
			\begin{align}
				H_{\text{rSSH}} & = \sum_{k = 0}^{N - 1} \left(\xi_k C_{\text{A}k}^\dagger C_{\text{B}k} + \xi^*_k C_{\text{B}k}^\dagger C_{\text{A}k}\right)\\
				& = \sum_{k = 0}^{N - 1} \mathbb{C}_k^\dagger \mathbb{H}_k \mathbb{C}_k,
			\end{align}
			where $\xi_k = 1 + g e^{i \theta_k}$, ${\mathbb{C}_k = \begin{pmatrix} C_{\text{A}k} & C_{\text{B}k}\end{pmatrix}^\intercal}$, and
			\begin{align}
				\mathbb{H}_k = \begin{pmatrix}
					0 & \xi_k\\
					\xi_k^* & 0
				\end{pmatrix}. \label{SSHHk}
			\end{align}

			In the thermodynamic limit, i.e., when $N \rightarrow \infty$ and $\theta_k$ can be treated as a continuous quantity in the range $[0, 2 \pi)$, the ground state and the ``first excited state'' become degenerate at the critical point $g = 1$. In another words, the two states share the same eigenenergy. Specifically, $\xi_k \rightarrow 0$ for values of $k$ such that $\theta_k \rightarrow \pi$ in Eq.~\eqref{SSHHk}. Therefore, the two corresponding eigenstates $C_{\text{A}k}^\dagger \ket{0}$ and $C_{\text{B}k}^\dagger \ket{0}$, where $\ket{0}$ is the vacuum state annihilated by all annihilation operators, form a basis for the degenerate subspace with zero eigenvalue.

			Next, we turn to the generator $K$ in the $g$-direction. From Eq.~\eqref{MainEquation}, together with Eq.~\eqref{Adiabatic}, we find that the generator $K$ for evolution along the $g$-direction can be expressed as
			\begin{align}
				K_\text{rSSH} = \sum_{k = 0}^{N - 1} \mathbb{C}_k^\dagger \mathbb{K}_k \mathbb{C}_k, \label{Kadi}
			\end{align}
			where
			\begin{align}
				\begin{split}
					\mathbb{K}_k & = \begin{pmatrix}
						\displaystyle \frac{- \sin \theta_k}{2 \left|\xi_k\right|^2} & \displaystyle t \frac{\cos \theta_k + g}{\xi_k^*}\\
						\displaystyle t \frac{\cos \theta_k + g}{\xi_k} & \displaystyle \frac{\sin \theta_k}{2 \left|\xi_k\right|^2}
					\end{pmatrix} + \begin{pmatrix}
						\alpha_k & \beta_k \xi_k\\
						\beta_k \xi_k & \alpha_k
					\end{pmatrix},
				\end{split}
			\end{align}

			with $\alpha_k$ and $\beta_k$ being arbitrary functions of $g$, corresponding to the residual gauge freedoms. For convenience, we set $\alpha_k = 0$ and $\beta_k = 0$ for all $k$, which simplifies the following discussion.

			When $\delta_k = \pi - \theta_k$ in Eq.~\eqref{Kadi} at $w = v$, $\mathbb{K}_k$ can be expanded as follows:
			\begin{align}
				\mathbb{K}_k = \frac{1}{2}\begin{pmatrix}
					\dfrac{-1}{\delta_k} + \dfrac{\delta_k}{12} & i \delta_k t\\
					- i \delta_k t & \dfrac{1}{\delta_k} - \dfrac{\delta_k}{12}
				\end{pmatrix} + \mathcal{O}\left(\delta_k^2\right).
			\end{align}

			Therefore, $K$ indeed becomes singular, since $\mathbb{K}_k$ becomes singular for some $k$ such that $\delta_k \rightarrow 0$.

			Nevertheless, as in the EP case, we provide another evolution generator
			\begin{align}
				K_\text{rSSH}' = \sum_{k = 0}^{N - 1} \mathbb{C}_k^\dagger \mathbb{K}'_k \mathbb{C}_k, \label{KrSSH1}
			\end{align}
			with
			\begin{align}
					\mathbb{K}'_k = \left\lbrace\begin{array}{ll}
					\begin{pmatrix}
						0 & - t\\
						- t & 0
					\end{pmatrix}, & \text{for $(\theta_k = \pi) \wedge (g = 1)$}\\
					\begin{pmatrix}
						\kappa^{(k)}_{11} & \kappa^{(k)}_{12}\\
						\kappa^{(k)}_{21} & \kappa^{(k)}_{22}
					\end{pmatrix}, & \text{otherwise}
				\end{array}\right., \label{KrSSH2}
			\end{align}
			where
			\begin{align}
				\kappa^{(k)}_{11} & = - \dfrac{\sin\theta_k \sin^2\left(2 t \left| \xi_k \right|\right)}{\left|\xi_k\right|^2} = - \kappa^{(k)}_{22},\\
				\begin{split}
					\kappa^{(k)}_{12} & = \dfrac{t (g + \cos \theta_k)}{\xi_k^*} + \dfrac{i \sin \theta_k \sin \left(2 t \left| \xi_k \right|\right)}{2 \xi_k^* \left| \xi_k \right|} = \kappa^{(k)*}_{21}.
				\end{split}
			\end{align}

			Note that
			\begin{align}
				\lim_{\theta_k \rightarrow \pi} \begin{pmatrix}
					\kappa^{(k)}_{11} & \kappa^{(k)}_{12}\\
					\kappa^{(k)}_{21} & \kappa^{(k)}_{22}
				\end{pmatrix} = \begin{pmatrix}
					0 & - t\\
					- t & 0
				\end{pmatrix},
			\end{align}
			when $g = 1$. Therefore, $K_\text{rSSH}'$, as defined in Eqs.~\eqref{KrSSH1} and \eqref{KrSSH2}, exhibits no singularity when $g = 1$, even at $\theta_k = \pi$.

			Similar to the EP case, the singularity at the DP that appears in the adiabatic gauge choice can be avoided by choosing a different gauge. Therefore, the information carried by the state remains unaffected.

			A formal discussion on how a DP can lead to an sigularity and a locally singularity-free choice of $K$ at a DP is provided in Appendix~\ref{App:DP}.

	\section{Physical Interpretations}

		Following the mathematical analysis, we find that the evolution generator of quantum states with respect to the parameter may exhibit singular behavior at EPs and DPs under the adiabatic gauge. This singularity, however, can be locally eliminated through an appropriate gauge choice.

		In the following, we discuss the underlying physics related to this effect.

		\subsection{Quantum Phase Transition}

			It is known that the eigenstate fidelity susceptibility is a useful tool for identifying the standard QPTs~\cite{Zanardi2006, Cejnar2007, Tzeng2008, Tzeng2008a, Sachdev2009, SHIJIAN2010}, the excited state QPTs~\cite{Cejnar2006, PerezFernandez2009}, and the EPs~\cite{Tu2023}, which is often related to QPT~\cite{Znojil2020, Znojil2021}. Not only do these QPTs occur when Hamiltonian eigenvalues become degenerate (i.e., at a DP or an EP), but their fidelity susceptibility also diverges at these points.

			In fact, this divergence is closely related to the aforementioned singularities that appear under the adiabatic gauge. To reveal the relation between the fidelity susceptibility and the emergent dimension evolution generator $K$, we begin with the eigenstate fidelity~\cite{Tzeng2021}, namely,
			\begin{align}
				\begin{split}
					& \mathcal{F} (\ket{\psi_n (q)}, \ket{\psi_n (q + \epsilon)})\\
					& = \Braket{\psi_n (q)}{\psi_n (q + \epsilon)} \Braket{\psi_n (q + \epsilon)}{\psi_n (q)},
				\end{split} \label{EigenFidelity}
			\end{align}
			where $\ket{\psi_n (q)}$ [$\ket{\psi_n (q + \epsilon)}$] is the $n$th eigenstate of $H(q)$ [$H(q + \epsilon)$] with $\Braket{\psi_m (q)}{\psi_n (q)} = \delta_{mn}$.

			Since we are comparing the fidelity between eigenstates at nearby different points, the adiabatic gauge must be applied.

			Using Eqs.~\eqref{qEvolution} and \eqref{Metricinq}, the fidelity shown in Eq.~\eqref{EigenFidelity} can be expanded as
			\begin{align}
				\mathcal{F} (\ket{\psi_n (q)}, \ket{\psi_n (q + \epsilon)}) = 1 - \epsilon^2 \chi_n(q) + \mathcal{O} \left( \epsilon^3 \right),
			\end{align}
			where the first order term in $\epsilon$ vanishes and $\chi_n$ is the fidelity susceptibility of the $n$th eigenstate~\cite{Ju2024}, which can be written as
			\begin{align}
				& \chi_n (q) = \langle K^2 \rangle_n (q) - \langle K \rangle_n^2 (q)\\
				& = \Bra{\psi_n (q)} K^2 (q) \Ket{\psi_n (q)} - \Bra{\psi_n (q)} K (q) \Ket{\psi_n (q)}^2.
			\end{align}

			As mentioned at the beginning of this subsection, QPTs occur when certain Hamiltonian eigenvalues become degenerate (either at a DP or an EP). Since we have already shown in the previous section that the $q$-evolution generator $K$ diverges at these points (see Fig.~\ref{LevelCrossing} for a schematic illustration), this singularity naturally leads to the divergence of the fidelity susceptibility $\chi_n$.

			\begin{figure}
				\begin{tikzpicture}
					\draw[->, gray, line width = 1] (-3, 0) -- (3, 0) node[right] {$q$};
					\draw[->, gray, line width = 1] (0, -1) -- (0, 3) node[above] {$E$};

					\draw[domain = -3: 3, smooth, variable=\x, color = black, line width = 1] plot ({\x}, {0.03 * (\x + 3) * (\x - 0.2 + 3) + 0.1});
					\draw[domain = -3: 3, smooth, variable=\x, color = black, line width = 1, densely dashed] plot ({\x}, {0.02 * (\x - 3) * (\x - 3) + 0.4});
					\draw[domain = -3: 3, smooth, variable=\x, color = black, line width = 1] plot ({\x}, {0.02 * (\x) * (\x) + 1.5});
					\draw[domain = -3: 3, smooth, variable=\x, color = black, line width = 1] plot ({\x}, {0.03 * (\x + 0.5) * (\x + 0.5) + 2});

					\node[anchor = west, color = black] at (3.1, 0.4) {$E'_{0/1}$};
					\node[anchor = west, color = black] at (3.1, 1.1) {$E_{0/1}$};
					\node[anchor = west, color = black] at (3.1, 1.7) {$E_2$};
					\node[anchor = west, color = black] at (3.1, 2.4) {$E_3$};

					\draw[color = black, dashed] (0.8, 3) -- (0.8, -0.2) node[below] {$q_0$};
				\end{tikzpicture}

				\caption{A schematic diagram of a QPT shown in an energy-level diagram. The horizontal axis represents the parameter $q$, and the vertical axis denotes the energy. The critical point of the QPT occurs at $q = q_0$, where the ground state and the first excited state exchange roles (i.e., an energy-level crossing). The same parameter $q_0$ also corresponds to the singularity of the $q$-evolution generator.}
				\label{LevelCrossing}
			\end{figure}

			As a consequence, the singularities in $K$ can be used to determine the critical points of QPTs (e.g., the singularity at $g = 1$ in the SSH example provided above). Hence, the singularities in the adiabatic gauge provide a hint about where QPTs occur experimentally.

		\subsection{(Reversible) Quantum Information Evolution and the Classical Black Hole Analogy}

			As mentioned earlier, the adiabatic gauge is physically appealing in many contexts; nevertheless, certain singularities can appear at EPs and DPs.

			On the other hand, we have also demonstrated that these singularities can be removed by choosing a different gauge. Hence, the singularities are gauge singularities, meaning that they are not physical but rather induced by the gauge choice. In other words, the information carried by a quantum state is not erased by the EPs or DPs (and therefore not by the QPTs). More accurately, the evolution of a quantum state passing through an EP or DP is reversible.

			In fact, this phenomenon is very similar to an object passing through the event horizon of a black hole.

			To see this, note that the evolution generators(both $H$ and $K$) are fiber-bundle generalizations of the Christoffel symbols in (pseudo-)Riemannian geometry. The singularities in $K$ that appear under the adiabatic gauge are, in fact, analogous to the singularities at the event horizon of classical black holes, where the singularities are coordinate singularities.

			As an example, the singularity at the event horizon of a Schwarzschild (stationary) black hole~\cite{Schwarzschild1916, Misner2017} disappears when a different coordinate system is chosen~\cite{Fromholz2014} (e.g., Eddington–Finkelstein coordinates~\cite{Eddington1924, Finkelstein1958} and Kruskal–Szekeres coordinates~\cite{Szekeres1959, Kruskal1960}).

			In addition, since the Ricci curvature tensor (and therefore the Ricci scalar) vanish at the event horizon, an object falling into the black hole does not experience any abrupt change when crossing the horizon.

			Therefore, QPTs are, to some extent, analogous to the event horizon of a classical black hole in many ways.

	\section{Conclusions}

		The behavior of quantum states around EPs and DPs (or the critical points of QPTs) is perplexing. In particular, many studies related to the Berry potential suggest that the Berry potentials are singular at these critical points. 

		To uncover the origin of these singularities, we trace them back to the evolution generator in the parameter-induced dimension and find that they are not physical singularities. Rather, they are mathematically similar to the singularities found at the event horizon of a classical black hole.

		In other words, these singularities can be removed, at least locally. Therefore, rather than being destroyed by these critical points, as the singular behavior of the Berry potential might suggest, the information carried by quantum states is preserved, as if the quantum states do not sense the presence of the critical points.

		Nevertheless, although the singularities at critical points can be locally removed, it does not mean the singularities lack physical meaning. Specifically, the singular property can manifest in a different way, like the relation between the event horizon and the temperature or entropy of a black hole.

		Therefore, we should not overlook the significance of these singularities.

	\begin{acknowledgments}
		The authors thank Prof. Chung-Hsien Chou at NCKU for inspiring this study, and Prof. Guang-Yin Chen and Prof. Chung-Yi Lin at NCHU for their helpful discussions. C.Y.J. is partially supported by the National Science and Technology Council through Grant No. NSTC 112-2112-M-110-013-MY3. 
	\end{acknowledgments}

	\begin{appendix}

		\section{Brief Review of Berry Potential \label{App:BerryPotential}}

			We give a brief review of the Berry potential and its gauge dependence in the Hermitian regime in this appendix. We also briefly comment on how this formalism can be generalized to the non-Hermitian regime.

			The adiabatic theorem~\cite{Born1928, Kato1950} states that if a Hamiltonian $H(t)$ varies slowly and has no degeneracies, a system starting in an eigenstate $\ket{\phi_n(t = 0)} = \ket{n (t = 0)}$ remains in the instantaneous eigenstate $\ket{n (t)}$ up to a phase factor, i.e.,
			\begin{align}
				\ket{\phi_n (t)} = e^{i \theta(t)} \ket{n (t)},
			\end{align}
			where $H(t) \ket{n (t)} = E(t) \ket{n (t)}$.

			Therefore, when the Hamiltonian depends on a parameter $q$, physically varying the parameter can be understood as making the parameter time-dependent, i.e., $q = q(t)$. This renders the corresponding Hamiltonian time-dependent, namely, $H = H(q(t))$. If the change in parameter is slow, the adiabatic theorem can also be applied here, so that
			\begin{align}
				\ket{\phi_n(t)} = e^{i \theta_n(t)} \ket{n(q(t))},
			\end{align}
			where $H(q(t)) \ket{n(q(t))} = E(q(t)) \ket{n(q(t))}$.

			This total phase $\theta(t)$ contains two parts: a dynamical phase $\displaystyle - \int_0^t dt' E_n(t')$ and a geometric (Berry) phase $\displaystyle \gamma_n(t)$. In other words, $\displaystyle \theta(t) = \gamma_n(t) - \int_0^t dt' E_n(t')$.

			Then the Schr\"{o}dinger equation for $\ket{\psi_n}$ becomes
			\begin{align}
				& i \frac{d}{dt} \ket{\phi_n(t)} = H(q(t)) \ket{\phi_n(q(t))}\\
				\Rightarrow & i \frac{d}{dt} \left( e^{i \theta_n(t)} \ket{n (q(t))} \right) = e^{i \theta_n(t)} H(q(t)) \ket{n(q(t))}\\
				\begin{split}
					\Rightarrow & - \left[\frac{d}{dt} \theta_n(t)\right] \left( e^{i \theta(t)} \ket{n(q(t))} \right) + i e^{i \theta_n(t)} \frac{d}{dt} \ket{n(q(t))}\\
					& \quad = e^{i \theta_n(t)} H(q(t)) \ket{n(q(t))}
				\end{split}\\
				\begin{split}
					\Rightarrow & \Bigg\{\left[E_n(t) - \frac{d}{dt} \gamma_n(t)\right] \left( e^{i \theta_n(q(t))} \ket{n (t)} \right)\\
					& \quad + i e^{i \theta_n(t)} \frac{d}{dt} \ket{n(q(t))}\Bigg\} = e^{i \theta_n(t)} E_n(q(t)) \ket{n(q(t))}
				\end{split}\\
				\Rightarrow & i \frac{d}{dt} \ket{n(q(t))} = \left[\frac{d}{dt} \gamma_n(t)\right] \ket{n(q(t))}\\
				\Rightarrow & i \frac{d q(t)}{dt} \partial_q \ket{n(q)}|_{q=q(t)} = \left[\frac{d}{dt} \gamma_n(t)\right] \ket{n(q(t))}. \label{TheSplittingPoint}
			\end{align}

			By taking the inner product with $\bra{n(q(t))}$, we obtain
			\begin{align}
				\begin{split}
					& \left[\frac{d}{dt} \gamma_n(t)\right] \braket{n(q(t))}{n(q(t))}\\
					& \quad = i \frac{d q(t)}{dt} \bra{n(q(t))}\partial_q \ket{n(q)}
				\end{split}\\
				\Rightarrow & \frac{d}{dt} \gamma_n(t) = i \frac{d q(t)}{dt} \bra{n(q(t))}\partial_q \ket{n(q)}\\
				\Rightarrow & \int_0^t dt' \frac{d}{dt'} \gamma_n(t') = i \int_0^t dt'\frac{d q(t')}{dt'} \bra{n(q(t'))}\partial_q \ket{n(q)}\\
				\begin{split}
					\Rightarrow & \gamma_n(t) - \gamma_n(0) = i \int_{q(0)}^{q(t)} dq \bra{n(q)}\partial_q \ket{n(q)}. \label{BerryPhaseTemp}
				\end{split}
			\end{align}

			In other words, the Berry phase does not depend on $t$ but rather on the initial parameter $q_0 = q(0)$ and final parameter $q = q(t)$.

			Since $\gamma_n(0) = 0$, by construction, Eq.~\eqref{BerryPhaseTemp} becomes
			\begin{align}
				\gamma_n (q) & = i \int_{q_0}^{q} dq' \bra{n(q')}\partial_{q'} \ket{n(q')}\\
				& = i \int_{q_0}^{q} dq' A_n(q'),
			\end{align}
			where $A_n(q) = \bra{n(q)}\partial_q \ket{n(q)}$ is the Berry potential.

			Note the slightly abuse of notation that $\gamma_n$ is now a function of $q$, since it does not really depend on $t$ but only the final parameter $q$.

			The eigenstate $\ket{\phi_n (t)}$ is defined up to a time-dependent phase factor, $\ket{\phi_n (t)} \rightarrow e^{i \chi_n(q(t))} \ket{\phi_n (t)}$, which changes the Berry connection
			\begin{align}
				\mathcal{A}_n = i \bra{n} \partial_q \ket{n} \to \mathcal{A}_n(t) - \partial_q \chi_n(q), \label{BerryPotentialGaugeTransform}
			\end{align}
			without chaning the physics. Therefore, there is a residual gauge degree of freedom in the choice of Berry potential.

			To generalize the discussion to the non-Hermitian regime, we simply need to replace the $\bra{n(q(t))}$ after Eq.~\eqref{TheSplittingPoint} with $\Bra{n(q(t))} = \bra{n(q(t))} G(t)$, where $G(t)$ transforms $\bra{n(q(t))}$ into the corresponding left eigenstate of $\ket{n(q(t))}$, such that $\Braket{m(q(t)}{n(q(t))}$. This is always possible if the system is not at an EP~\cite{Ju2019}.

			As a result, the Berry potential in the non-Hermitian regime becomes
			\begin{align}
				A_n(q) = \Bra{n(q)} \partial_q \Ket{n(q)},
			\end{align}
			obtained simply by replacing $\bra{n(q)}$ with the left eigenstate corresponding to $\ket{n(q)}$.

		\section{The Residual and Berry Potential Gauge Freedoms \label{App:GaugeFreedoms}}

			As mentioned at the end of Appendix~\ref{App:BerryPotential}, the Hamiltonian eigenstates are not uniquely defined.

			Since $\ket{\widetilde{\phi}_{n}(q)} = e^{-i \chi_{n}(q)} \ket{\phi_{n}(q)}$ is still an eigenstate for any function $\chi_{n}(q)$, the corresponding Berry potential for the eigenstate $\ket{\widetilde{\phi}_n(q)}$ is
			\begin{align}
				A'_{n}(q) & = i \Bra{\widetilde{\phi}_{n} (q)} \left(\partial_q \Ket{\widetilde{\phi}_{n} (q)}\right)\\
				& = i \Bra{\phi_{n} (q)} e^{i \chi_{n}(q)} \left(\partial_q e^{- i \chi_{n}(q)} \Ket{\phi_{n} (q)}\right)\\
				& = i \Bra{\phi_{n} (q)} \left(\partial_q \Ket{\phi_{n} (q)}\right) + \partial_q \chi_{n}(q),
			\end{align}
			which is a gauge-transformed Berry potential from $A_{n}(q)$.

			This is, in fact, the standard Berry potential transformation shown in Eq.~\eqref{BerryPotentialGaugeTransform}.

			This gauge transformation is consistent with the the residual gauge transformation in Eq.~\eqref{ResidualGaugeTransformation}. Specifically, the Berry potential corresponding to the gauge-transformed $K' = K + \Delta \! K$ is
			\begin{align}
				A'_{n}(q) & = \Bra{\psi_{n} (0, q)} \left(K^{(0)} + \Delta \! K \right) \Ket{\psi_{n} (0, q)}\\
				\begin{split}
					& = \Bra{\psi_{n} (0, q)} K^{(0)} \Ket{\psi_{n} (0, q)}\\
					& \quad + \Bra{\psi_{n} (0, q)} \Delta \! K \Ket{\psi_{n} (0, q)}
				\end{split}.
			\end{align}
			Since $[\Delta \! K, H] = 0$, $\Delta \! K$ shares the same eigenstates as $H$. We can therefore choose the eigenvalue of $\Delta \! K$ correspdoning to the eigenstate $\ket{\psi_{n} (0, q)}$ to be $\partial_q \chi_{n}(q)$ [see Eq.~\eqref{DeltaKEigenvalue}], so that
			\begin{align}
				A'_{n}(q) & = A_{n}(q) + \Bra{\psi_{n} (0, q)} \partial_q \chi_{n}(q) \Ket{\psi_{n} (0, q)}\\
				& = A_{n}(q) + \partial_q \chi_{n}(q).
			\end{align}

			We can also obtain the Berry potential by defining the eigenstate as $\ket{\phi_{n}(q)} = \ket{\psi_{n}(t_0, q)} = e^{-i h_{n}(q) t_0} \ket{\psi_{n}(t = 0, q)}$, i.e., at any fixed time $t = t_0$. The Berry potentials are the same up to an gauge transformation, by identifying $\chi_{n}(q) = h_{n}(q) t_0$.

		\section{The Adiabatic Gauge and Eigenstate Evolution \label{App:Adiabatic}}
			From the adiabatic gauge,
			\begin{align}
				\left[\partial_t K, H\right] = 0,
			\end{align}
			we know that $\partial_t K$ and $H$ share a set of eigenstates. Let $\{\ket{\psi_i}\}$ be the set of eigenstates shared by both $H$ and $\partial_t K$, i.e.,
			\begin{align}
				\left\{\begin{array}{l}
					H \ket{\psi_i} = h_i \ket{\psi_i}\\
					\left(\partial_t K\right) \ket{\psi_i} = k'_i \ket{\psi_i}
				\end{array}\right. ,
			\end{align}
			the corresponding eigenvalues of $\partial_t K$ can be determined by Eqs.~\eqref{qEvolution} and \eqref{MainEquation}. To be more specific,
			\begin{align}
				k'_i \ket{\psi_i} & = \left(\partial_t K\right) \ket{\psi_i}\\
				& = \left(i \left[K, H\right] + \partial_q H\right) \ket{\psi_i}\\
				& = i K H \ket{\psi_i} - i H K \ket{\psi_i} + \left(\partial_q H\right) \ket{\psi_i}\\
				& = i h_i K \ket{\psi_i} - i H K \ket{\psi_i} + \left(\partial_q H\right) \ket{\psi_i}\\
				& = - h_i \partial_q \ket{\psi_i} + H \partial_q \ket{\psi_i} + \left(\partial_q H\right) \ket{\psi_i}\\
				& = - h_i \partial_q \ket{\psi_i} + \partial_q \left(H \ket{\psi_i}\right)\\
				& = - h_i \partial_q \ket{\psi_i} + \partial_q \left(h_i \ket{\psi_i}\right)\\
				& = \left(\partial_q h_i\right) \ket{\psi_i}.
			\end{align}
			In other words, the eigenvalue of $\partial_t K$ corresponding to the eigenstate $\ket{\psi_i}$ is $k'_i = \partial_q h_i$.

			Therefore, we conclude that
			\begin{align}
				\left[\partial_t K, H\right] = 0 \quad \Leftrightarrow \quad \left( \partial_t K \right) \ket{\psi_i} = \left( \partial_q h_i \right) \ket{\psi_i},
			\end{align}
			where $h_i$ and $\ket{\psi_i}$ are the eigenvalues and eigenstates of the Hamiltonian $H$, i.e., $H \ket{\psi_i} = h_i \ket{\psi_i}$.

		\section{Example --- From $K$ to $A_i$ \label{App:ExampleKToA}}
			In this appendix, we provide a concrete example of how to obtain the Berry potential from the parameter-induced dimension generator. Specifically, we consider a non-Hermitian Hamiltonian that depends on a parameter $z$ and exhibits an EP at $z = 0$.

			We take one of the Hamiltonians presented in \cite{MehriDehnavi2008}, namely,
			\begin{align}
				H(z) = \begin{pmatrix}
					z & 1\\
					0 & - z
				\end{pmatrix},
			\end{align}
			where $z$ is the parameter, and find the corresponding Berry potential.

			Given this Hamiltonian, the (right) eigenstates are
			\begin{align}
				\Ket{\phi_1} = \begin{pmatrix}
					1\\
					0
				\end{pmatrix}, \quad \Ket{\phi_2} = \begin{pmatrix}
					- 2 z\\
					1
				\end{pmatrix},
			\end{align}
			while the dual eigenstates that satisfy $\Braket{\phi_i}{\phi_j} = \delta_{ij}$ (i.e., the left eigenstates) are
			\begin{align}
				\Bra{\phi_1} = \begin{pmatrix}
					1 & 2 z
				\end{pmatrix}, \quad \Bra{\phi_2} = \begin{pmatrix}
					0 & 1
				\end{pmatrix}.
			\end{align}

			The $z$-generator $K$, in the adiabatic gauge, is given by
			\begin{align}
				K(t, z) = \begin{pmatrix}
					t & \frac{t}{z}\\
					0 & - t + \frac{i}{z}
				\end{pmatrix},
			\end{align}
			up to some residual gauge freedom.

			The Berry potentials for both eigenstates can then be computed as
			\begin{align}
				A_1 = \Bra{\phi_1} K(t = 0, q) \Ket{\phi_1} = 0\\
				A_2 = \Bra{\phi_2} K(t = 0, q) \Ket{\phi_2} = \frac{i}{z},
			\end{align}
			which are consistent with the results in \cite{MehriDehnavi2008}.

			Notice that the Berry potential for the eigenstate $\ket{\phi_2}$ is singular at the EP (i.e., $z = 0$). This singularity is, in fact, characteristic of an EP.

		\section{In the Vicinity of an Exceptional Point \label{App:EP}}

			We argued in the main text that the singularity at an EP comes from the incompatibility between the adiabatic gauge and the non-diagonalizability of the Hamiltonian at the EP. Nevertheless, we provide a different choice of $K$ that does not lead to a singularity at the EP.

			Although the Hamiltonian $H_{\text{\tiny EP}}(q)$ is not diagonalizable at $q = q_{\text{\tiny EP}}$, we can still make a similar transformation from the original Hamiltonian at the EP to a Jordan form, i.e.,
			\begin{align}
				J_\text{\tiny EP} = Q_{\text{\tiny EP}}^{-1} H_{\text{\tiny EP}}(q_{\text{\tiny EP}}) Q_{\text{\tiny EP}}, \label{EPBlock}
			\end{align}
			where $J_\text{\tiny EP}$ is a matrix of Jordan form. In other words,
			\begin{align}
				\setlength{\arraycolsep}{0pt}
				J_\text{\tiny EP} = \begin{pmatrix}
					\boxed{J_1} & 0 & \cdots\\
					0 & \boxed{J_2} & \cdots\\
					\vdots & \vdots & \ddots
				\end{pmatrix},
			\end{align}
			where each $J_i$ is a Jordan block,
			\begin{align}
				J_i = \begin{pmatrix}
					\lambda_i & c_i & 0 & \cdots\\
					0 & \lambda_i & c_i & \cdots\\
					0 & 0 & \lambda_i & \cdots\\
					\vdots & \vdots & \vdots & \ddots
				\end{pmatrix},
			\end{align}
			with $c_i$ being an arbitrarily nonzero constant with the same unit as the Hamiltonian.

			To keep the discussion simple, we focus on the case where all the Hamiltonian eigenstates coalesce into one (the argument can easily be extended to more general cases through block diagonalization), namely,
			\begin{align}
				J_\text{\tiny EP} = \begin{pmatrix}
					\lambda & c & 0 & \cdots\\
					0 & \lambda & c & \cdots\\
					0 & 0 & \lambda & \cdots\\
					\vdots & \vdots & \vdots & \ddots
				\end{pmatrix},
			\end{align}
			where $c$ is a nonzero constant with the same unit as the Hamiltonian.

			Although $H_{\text{\tiny EP}}(q)$ can only be put in the Jordan form at $q = q_{\text{\tiny EP}}$, when $q \neq q_{\text{\tiny EP}}$, $H_{\text{\tiny EP}}(q)$ can still be diagonalized as
			\begin{align}
				\widetilde{\Lambda}(q) = \widetilde{P}^{-1}(q) H_{\text{\tiny EP}}(q) \widetilde{P}(q), \label{ToBeBlockDiag}
			\end{align}
			where $\widetilde{\Lambda}(q) = \text{diag}\left( \lambda_1(q) , \lambda_2(q), \cdots \right)$.

			To show that the evolution generator $K$ can be made continuous at $q = q_{\text{\tiny EP}}$, we modify Eq.~\eqref{ToBeBlockDiag} to
			\begin{align}
				\widetilde{J}(q) = \widetilde{S}^{-1}(q) \widetilde{P}^{-1}(q) H_{\text{\tiny EP}}(q) \widetilde{P}(q) \widetilde{S}(q),
			\end{align}
			where
			\begin{align}
				\widetilde{J}(q) = \begin{pmatrix}
					\lambda_1(q) & c & 0 & \cdots\\
					0 & \lambda_2(q) & c & \cdots\\
					0 & 0 & \lambda_3(q) & \cdots\\
					\vdots & \vdots & \vdots & \ddots
				\end{pmatrix},
			\end{align}
			by choosing
			\begin{align}
				\widetilde{S} = \begin{pmatrix}
					1 & \dfrac{c}{\lambda_{12}} & \dfrac{c^2}{\lambda_{12} \lambda_{13}} & \cdots\\
					0 & \dfrac{c}{\lambda_{21}} & \dfrac{c^2}{\lambda_{21} \lambda_{23}} & \cdots\\
					0 & 0 & \dfrac{c^2}{\lambda_{31} \lambda_{32}} & \cdots\\
					\vdots & \vdots & \vdots & \ddots
				\end{pmatrix},
			\end{align}
			where $\lambda_{ij} \equiv \lambda_i - \lambda_j$.

			For conciseness, we define ${\widetilde{Q}(q) \equiv \tilde{P}(q) \widetilde{S}(q)}$. Moreover, since $\widetilde{P}$ is not uniquely defined, we can always rescale the columns of $\widetilde{P}$ so that $\displaystyle Q_{\text{\tiny EP}} = \lim_{q \rightarrow q_{\text{\tiny EP}}} \widetilde{Q}(q)$.

			Letting $K = R^{-1} F R$, where $R$ satisfies
			\begin{align}
				\partial_t R = i R H_{\text{\tiny EP}}, \label{REq}
			\end{align}
			so that
			\begin{align}
				F = \int dt R \left( \partial_q H_{\text{\tiny EP}} \right) R^{-1}.
			\end{align}

			By choosing $R(t = 0) = \mathds{1}$, we find
			\begin{align}
				R = \widetilde{Q}^{-1} \widetilde{W} \widetilde{Q},
			\end{align}
			where
			\begin{align}
				\widetilde{W} = \begin{pmatrix}
					e^{i \lambda_1 t} & c \dfrac{E_{12}}{\lambda_{12}} & \dfrac{c^2}{\lambda_{12}} \left( \dfrac{E_{13}}{\lambda_{13}} - \dfrac{E_{23}}{\lambda_{23}} \right) & \cdots\\[10pt]
					0 & e^{i \lambda_2 t} & c \dfrac{E_{23}}{\lambda_{23}} & \cdots\\[10pt]
					0 & 0 & e^{i \lambda_3 t}\\
					\vdots & \vdots & \vdots & \ddots
				\end{pmatrix},
			\end{align}
			and $E_{ij} \equiv \exp(i \lambda_{i} t) - \exp(i \lambda_{j} t)$. It is clear that when $q$ approaches $q_{\text{\tiny EP}}$, all the $E_{ij}$ and $\lambda_{ij}$ become closer to zero. When $q \rightarrow q_{\text{\tiny EP}}$, the limit of $\widetilde{W}$ becomes
			\begin{align}
				W_\text{\tiny EP} = \lim_{q \rightarrow q_{\tiny EP}} \widetilde{W} = e^{i \lambda t} \begin{pmatrix}
					1 & i c t & \dfrac{- c^2 t^2}{2}& \cdots\\
					0 & 1 & i c t& \cdots\\
					0 & 0 & 1& \cdots\\
					\vdots & \vdots & \vdots & \ddots
				\end{pmatrix}, \label{WEP}
			\end{align}
			such that
			\begin{align}
				W_\text{\tiny EP}^{-1} \partial_t W_\text{\tiny EP} = i J_\text{\tiny EP}.
			\end{align}		
			Hence, we can define
			\begin{align}
				Q = \left\lbrace \begin{array}{ll}
					\widetilde{Q}(q), & \text{for }	q \neq q_{\text{\tiny EP}}\\
					Q_{\text{\tiny EP}}, & \text{for } q = q_{\text{\tiny EP}}
				\end{array}\right.,\\
				W = \left\lbrace \begin{array}{ll}
					\widetilde{W}(q), & \text{for }	q \neq q_{\text{\tiny EP}}\\
					W_\text{\tiny EP}, & \text{for } q = q_{\text{\tiny EP}}
				\end{array}\right.,
			\end{align}
			so that both $Q$ and $W$ are continuous in the vicinity of $q = q_{\text{\tiny EP}}$.

			Since $\partial_q H$, by construction, is continuous everywhere,
			\begin{align}
				K & = Q^{-1} W^{-1} \left[ \int_{t_0}^t dt W Q^{-1} \left( \partial_q H_{\text{\tiny EP}} \right) Q W^{-1} \right] W Q,\\
				& = Q^{-1} W^{-1} \left[ \mathcal{I}(t) - \mathcal{I}(t_0) \right] W Q
			\end{align}
			where $\partial_t \mathcal{I} = W Q^{-1} \left( \partial_q H_{\text{\tiny EP}} \right) Q W^{-1}$ and the $\mathcal{I}(t_0)$ is a gauge choice term. Since $W Q^{-1} \left( \partial_q H_{\text{\tiny EP}} \right) Q W^{-1}$ is continuous in the vicinity, $K$ is also continuous in the vicinity at $q = q_{\text{\tiny EP}}$.

			Note that $K$ contains terms with higher power of $t$ because $\mathcal{I}(t)$ generally has terms with higher power in $t$ and exponentials of t [see Eq.~\eqref{WEP}]. Therefore, the evolution generator $K$ is not compatible with the adiabatic gauge.

		\section{In the Vicinity of a Diabolic Point \label{App:DP}}

			In this section, we study the case when the eigenvalues of $H_{\text{\tiny DP}}(q)$ degenerate at $q = q_{\text{\tiny DP}}$, i.e., when some eigenvalues of $H_{\text{\tiny DP}}(q_{\text{\tiny DP}})$ become identical, while $H_{\text{\tiny DP}}(q_{\text{\tiny DP}})$ remains diagonalizable.

			We first introduce an invertible matrix $P(q)$ and a diagonal matrix $\Lambda(q) = \text{diag}(\lambda_1(q), \lambda_2(q), \cdots)$ that conjugates to $H_{\text{\tiny DP}}(q)$, such that 
			\begin{align}
				\Lambda(q) = P^{-1}(q) H_{\text{\tiny DP}}(q) P(q), \label{DiagH}
			\end{align}
			where the $\lambda$'s are the eigenvalues of $H_{\text{\tiny DP}}$.

			We then express $K$ as $K = R^{-1} F R$, where
			\begin{align}
				\partial_t R = i R H_{\text{\tiny DP}},
			\end{align}
			so that $F$ satisfies
			\begin{align}
				& \partial_t F = R \left( \partial_q H_{\text{\tiny DP}} \right) R^{-1}\\
				\Rightarrow & F = \int dt R \left( \partial_q H_{\text{\tiny DP}} \right) R^{-1}. \label{FEquation}
			\end{align}
			Without loss of generality, we choose ${R(t = 0) = \mathds{1}}$, which gives
			\begin{align}
				R(t) = P^{-1} D P,
			\end{align}
			where
			\begin{align}
				D = \text{diag}\left( e^{i \lambda_1 t}, e^{i \lambda_2 t}, \cdots \right).
			\end{align}
			Thus, Eq.~\eqref{FEquation} becomes 
			\begin{align}
				F = P^{-1} \left[\int dt D P \left( \partial_q H_{\text{\tiny DP}} \right) P^{-1} D^{-1} \right] P.
			\end{align}
			Since, by construction, $\partial_q H_{\text{\tiny DP}}$ is nonsingular everywhere and $P$ can be chosen that both $P$ and $P^{-1}$ are continuous in the vicinity of $q = q_{\text{\tiny DP}}$, we define
			\begin{align}
				M = P \left( \partial_q H_{\text{\tiny DP}} \right) P^{-1} = \begin{pmatrix}
					m_{11} & m_{12} & \cdots\\
					m_{21} & m_{22} & \cdots\\
					\vdots & \vdots & \ddots
				\end{pmatrix},
			\end{align}
			which is also continuous around $q = q_{\text{\tiny DP}}$. Finally, we have 
			\begin{align}
				K = P^{-1} D^{-1} \left[\int dt D M D^{-1} \right] D P.
			\end{align}

			A straightforward calculation shows that
			\begin{align}
				\begin{split}
					& \left( P K P^{-1} \right)_{ij} = \left\lbrace D^{-1} \left[ \int dt D M D^{-1} \right] D \right\rbrace_{ij}\\
					& \quad = e^{- i \left( \lambda_i - \lambda_j \right) t} \int dt e^{i \left( \lambda_i - \lambda_j \right) t} m_{ij}\\
					& \quad = \left\lbrace \begin{array}{l l}
						m_{ij} t + \Delta \! K_{ij}, & \text{for } \lambda_i = \lambda_j\\
						\dfrac{- i m_{ij}}{\lambda_i - \lambda_j} + \Delta \! \widetilde{K}_{ij} e^{- i \left( \lambda_i - \lambda_j \right) t}, & \text{for } \lambda_i \neq \lambda_j
					\end{array}\right.,
				\end{split} \label{DPSingular}
			\end{align}
			where $\Delta \! K_{ij}$ and $\Delta \! \widetilde{K}_{ij}$ are arbitrary time-independent matrix elements that correspond to the gauge choice.

			Naively, because at least two $\lambda$ values can be arbitrarily close to each other, the denominator $\lambda_i - \lambda_j$ approaches zero as $q \rightarrow q_{\text{\tiny DP}}$, making Eq.~\eqref{DPSingular} appear to diverge at $q = q_{\text{\tiny DP}}$. Nevertheless, we can choose
			\begin{align}
				\Delta \! K_{ij} = 0 \quad \text{and} \quad\Delta \! \widetilde{K}_{ij} = \frac{i m_{ij}}{\lambda_i - \lambda_j},
			\end{align}
			so that
			\begin{align}
				\left( P K P^{-1} \right)_{ij} = \left\lbrace \begin{array}{l l}
						m_{ij} t, & \text{for } \lambda_i = \lambda_j\\~\\
						\dfrac{ i m_{ij}}{\lambda_{j} - \lambda_{i}}\left[ 1 - e^{ i \left( \lambda_{j} - \lambda_{i} \right) t} \right], & \text{for } \lambda_i \neq \lambda_j
					\end{array}\right..
			\end{align}
			To show that $PKP^{-1}$ can be continuous in the vicinity of $q = q_{\text{\tiny DP}}$, we take the limit $q \rightarrow q_{\text{\tiny DP}}$; as a result, $\lambda_i$ also approaches $\lambda_j$. Therefore,
			\begin{align}
				\begin{split}
					& \lim_{\lambda_i \rightarrow \lambda_j} \dfrac{ i m_{ij}}{\lambda_{j} - \lambda_{i}}\left[ 1 - e^{ i \left( \lambda_{j} - \lambda_{i} \right) t} \right]\\
					= & \lim_{\lambda_i \rightarrow \lambda_j} \frac{ i m_{ij}}{\lambda_{j} - \lambda_{i}}\left[ e^{0 \cdot t} - e^{ i \left( \lambda_{j} - \lambda_{i} \right) t} \right]\\
					= & ~ m_{ij} t.
				\end{split}
			\end{align}

			Hence, $P K P^{-1}$ is continuous at $q = q_{\text{\tiny DP}}$, which renders $K$ finite at $q = q_{\text{\tiny DP}}$.

			Nevertheless, since the components of $K$ can at most be linear in $t$ if the adiabatic gauge is applied, the $\Delta \! \widetilde{K}_{ij}$ in Eq.~\eqref{DPSingular} can only be zero. This implies that the limit diverges when the Hamiltonian at a DP if the corresponding $m_{ij} \neq 0$. Therefore, the evolution generator $K$ is singular at a DP under the adiabatic gauge.

	\end{appendix}

	\bibliography{References}

\end{document}